# CAPITAL STRUCTURE AND SPEED OF ADJUSTMENT IN U.S. FIRMS. A COMPARATIVE STUDY IN MICROECONOMIC AND MACROECONOMIC CONDITIONS-A QUANTILLE REGRESSION APPROACH.

**Andreas Kaloudis[1], Dimitrios Tsolis[2]**

[1] PhD Candidate, Cultural Heritage and New Technologies Department, University of Patras, antreaskaloudis25@gmail.com.

[2] Assistant Professor, Cultural Heritage and New Technologies Department, University of Patras, dtsolis@upatras.gr.

**ABSTRACT**

The major perspective of this paper is to provide more evidence regarding how "quickly", in different macroeconomic states, companies adjust their capital structure to their leverage targets. This study extends the empirical research on the topic of capital structure by focusing on a quantile regression method to investigate the behavior of firm-specific characteristics and macroeconomic factors across all quantiles of distribution of leverage (book leverage and market leverage). Therefore, depending on a partial adjustment model, we find that the adjustment speed fluctuated in different stages of book versus market leverage. Furthermore, while macroeconomic states change, we detect clear differentiations of the contribution and the effects of the firm-specific and the macroeconomic variables between market leverage and book leverage debt ratios. Consequently, we deduce that across different macroeconomic states the nature and maturity of borrowing influence the persistence and endurance of the relation between determinants and borrowing.

**INTRODUCTION**

Plausible questions have been triggered in the scientific area of capital structure dynamic determination prompted by the recent global financial crisis, regarding how "quickly", in different macroeconomic states, companies adjust their capital structure to their book leverage targets. In an effort to broaden the debate scope, we focus on SMEs and discuss the relative importance of macroeconomic and firm-specific variables, in changing macroeconomic conditions. Therefore, depending on a partial adjustment model, we obtain that book and market debt follow different patterns concerning their adjustment speeds.

A scientific area that has drawn research interest during the last two decades is capital structure determination of Small and Medium Enterprises (SMEs) and the reason is partly the recognition of the importance of SMEs for the economy in terms of employment and value added as well as numbers of enterprises and partly the acknowledgement that SMEs financing exhibits considerable differences compared to large enterprises. Hence, during the last twenty years, a series of research has emerged which focuses on investigating the particularities of small enterprises in their capital structure determination.

Particularly, this issue is tackled in Torres and Julien (2005) research from a managerial perspective, where they describe the main findings of researchers over the last three decades, which have led to the recognition of SMEs specificities. Thus, in this wide recognition where research has shed light on the fact that large firms' theory has limited applicability to SMEs, the theory of capital structure cannot be an exemption. Ang (1991) was the first that highlighted this approach by pointing out that the theory of finance was not developed with the small business in mind, whereas Cressy and Olofsson (1997) declare that small businesses are not scaled-down versions of large businesses and Michaelas et al. (1999) endeavors to relate the different theoretical attributes to small enterprises.

On the contrary, in his research Hackbarth et al. (2006) indicated that little attention has been paid to the macroeconomic conditions' effects on capital structure choices and credit risk, notwithstanding the substantial development of the capital structure literature. Plausible questions have been triggered in the scientific area of capital structure determination by the recent global financial crisis regarding how quickly firms tend to adjust their capital structure in different economic states to their long-term targets. Furthermore, the research of Cook and Tang (2010) is built on the findings of previous analysts, such as Chloe et al. 1993; Gertler and Gilchrist, 1993; Korajczyk and Levy, 2003, where macroeconomic conditions do affect companies' financing choices, and denote that companies adjust their leverage toward target faster in good macroeconomic states compared to bad macroeconomic states. In their study, Oztekin and Flannery (2012) compare speed adjustments of capital structure across countries and prove that financial and legal conditions vary with debt adjustment speeds. Correspondingly, Baum et al. (2016) in his research follow a similar approach and indicate that companies with above-target leverage and financial surpluses adjust their leverage more rapidly when macroeconomic risk is high and firm-specific risk is low, whilst companies with below-target leverage and financial deficits adjust their capital structure more quickly when both types of risk are low. The context of Baum et al. (2016) triggers precisely the main idea of this paper, namely to investigate in changing macroeconomic states the relative significance of traditional firm-specific capital structure determinants, such as asset structure, size, profitability, risk and growth, versus macroeconomic variables.

Contributing to the interesting debate of the relative importance of firm-specific versus macroeconomic variables in changing macroeconomic conditions is the main research objective of this paper, seen respectively in a demand-driven context vs. supply-driven context, in the SMEs environment and none of the aforementioned studies simultaneously combine SMEs' specificities with adjustment speed in capital structure determination or explore the issue of firm-specific vs. macroeconomic variables in different economic states.

Two periods of different macroeconomic states are identified, growth and recession, that mainly follow the methodological explanation of Cook and Tang (2010) and Oztekin and Flannery (2012), by using a dynamic model of partial adjustment capital structure with unobserved heterogeneity and fake variable for the macroeconomic states. We find that there are clear differentiations of the effects and the contribution of the firm-specific versus the macroeconomic variables between long-term and short-term debt ratios in changing macroeconomic states. From our results we reach a conclusion that the nature and maturity of borrowing across different macroeconomic states affects the endurance and persistence of the relation between determinants and borrowing.

Thus, our opinion is that this paper contributes on the recent dynamic determination of capital structure in the below ways. Firstly, we manage to broaden the scope of the debate by including SMEs, secondly, we show that macroeconomic states have a prevailing effect on how the relationships of capital structure determinants and leverage are shaped and finally we indicate that these relationships are also influenced by debt maturity.

**METHODOLOGY AND DATA**

Initially, we probe if our panel data model is a random or fixed effect model. This can be detected through Hausman (1978) test, which allows us to find that our model is a fixed effect model. Generally, panel data models provide us control of the implications of companies' non-observable individual effects on the estimated parameters. To the extent of our knowledge panel data are the most appropriate to examine a dynamic phenomenon which varies across time compared to cross-section or time series data that neither express dynamic relations. Moreover, panel data allow us to estimate raised accuracy since they use double observations which are used in both assessment with the cross section or time series data.

**THE MODEL**

Following the rationale of Cook and Tang (2010), Oztekin and Flannery (2012), Nikolaos Daskalakis, Dimitrios Balios and Violetta Dalla (2017), Antreas Kaloudis and Dimitrios Tsolis (2018), we use a partial adjustment model, which assumes that the target debt ratio LEV $*_{i,t}$ from firm i at time t, is given by:

$$\text{LEVB} *_{i,t} = a * + a *_i + \beta * X_{i,t} + \gamma * M_t, \quad i = 1, \ldots, N, \quad t = 2, \ldots, T_i$$

$$\text{LEVM} *_{i,t} = a * + a *_i + \beta * X_{i,t} + \gamma * M_t, \quad i = 1, \ldots, N, \quad t = 2, \ldots, T_i$$

Where $a*$ is the constant term, $a*_i$ is the unobserved heterogeneity of firm i, $X = (X_1, \ldots, X_K)'$ and $M = (M_1, \ldots, M_J)'$ are (column) vectors of firm specific and macroeconomic variables respectively, $\beta* = (\beta*_1, \ldots, \beta*_K)$ is the (row) coefficient vector of firm-specific variables and $\gamma* = (\gamma*_1, \ldots, \gamma*_j)$ the (row) coefficient of the macroeconomic variables. The debt ratio $DR_{i,t}$ adjust to its target according to the rule:

$$DR_{i,t} - DR_{i,t-1} = \delta * (DR*_{i,t} - DR_{i,t-1}) + \varepsilon_{i,t}$$

$$LEVB_{i,t} - LEVB_{i,t-1} = \delta * (LEVB*_{i,t} - LEVB_{i,t-1}) + \varepsilon_{i,t}$$

$$LEVM_{i,t} - LEVM_{i,t-1} = \delta * (LEVM*_{i,t} - LEVM_{i,t-1}) + \varepsilon_{i,t}$$

**Where** $\delta *$ is the speed of adjustment and $\varepsilon_{i,t}$ is the error term.

**The Data**

The data is collected from published financial statements of U.S. economy companies for 44 years. We use a dataset of SME's and MNC's of United States economy. For the best of our knowledge panel data are the most appropriate to observe a dynamic phenomenon that variates cross time in comparison with cross-section and time series data which neither express dynamic relations nor produced estimates are highly accurate due to the multicollinearity existence. Furthermore, panel data provide us estimates of raised accuracy while they used more than the double number of total observations that is used in both assessment with the times series or cross section data

**The variables**

  **PROXIES FOR LEVERAGE**

Book leverage is the ratio of book debt/total assets.
Market leverage is the ratio of book debt / ( book debt + market equity )

**Firm specific factors**

  **LIQUIDITY (LIQTA)** is the ratio of current assets to current liabilities, as used by Graham and Harvey (2001) and De Jong et al. (2008).

  **NDTS** is the nondebt tax shields (tax shields excluding interest), as used by Titman and Wessels (1998), Barton et al. (1989), Prowse (1990) and Miguel and Pindado (2001), who found an inverse relationship between leverage and non-debt tax shields;
$NDTS_{i,t} = EBIT_{i,t} - IP_{i,t} - (\frac{T_{i,t}}{T})$,

Where $IP_{i,t}$ is the interest payable, $T_{i,t}$ is the income Tax, T is the corporate tax rate.

**PROFITABILITY (PROFTA).** In view of the pecking order theory, firm's financing decisions follow in general a hierarchy, preferring debt over equity and internal over external financing (Michaelas et al., 1999, Daskalakis & Psillaki, 2008; Psillaki & Daskalakis, 2009). Thus, it is expected that profitability should be negatively related to debt and be measured as earnings before interest and taxes to total assets

**SIZE (SIZEAT),** expected to be positively correlated with debt levels. Larger firms may be able to reduce the transaction costs associated with long-term debt issuance. Public corporate debt usually trades in large blocks relative to the size of an equity trade, and most issues are at least 100 million dollars in face value to provide liquidity. Marsh's (1982) survey concludes that large firms more often choose long-term debt while small firms choose short-term debt. Size is measured as the natural logarithm of total sales.

**GROWTH (GROWTHAT)** and leverage relation can be either negative or positive, with GR being measured as the annual rate of change in sales.

**INVESTMENTS (INVTA)** In accordance with Lewellen and Badrinath (1997), where: $I_{i,t} = NPPE_{i,t} - NPPE_{i,t-1} + D_{i,t}$, where $NPPE_{i,t}$ is the fixed assets and $D_{i,t}$ is a proxy for the depreciation.

The **market-to-book (MRBRATIO)** ratio was used by Rajan and Zingales (1995), De Jong et al. (2008), Lemmon and Zender (2010) and Sinan (2010).

**Macroeconomic factors**

**INFLATION**

An extensively investigated macroeconomic factor is the inflation rate (INFL). However, contradictory evidence exists concerning the effect of inflation on capital structure. In the context of literature, Bastos et al., (2009) found no effect of inflation on leverage, while Frank & Goyal (2009) detected a positive relationship between market leverage and inflation, yet no relationship on book leverage. On the other hand, Hanousek & Shyamshur (2011) verified that inflation generally has a positive influence on leverage, this effect however turns unimportant for certain specifications of their model. INFL is referred to the annual rate of change of the CPI index.

**GDP_RATE**

GDPgrowth indicates monetary conditions in general. Beck et al. (2008), De Jong et al. (2008), Chipeta & Mbululu (2013) and Muthama et al. (2013) discovered that firms that operate in a country with increased real GDP, have a higher level of economic wealth and therefore they tend to issue more debt than equity. On the other hand, Kayo & Kimura (2011) confirmed a negative relationship and argued that companies tend to generate higher net incomes and greater revenues during periods of peak economic activity. In view of this, the opportunity to finance further investments internally and not by issuing equity or debt is provided.

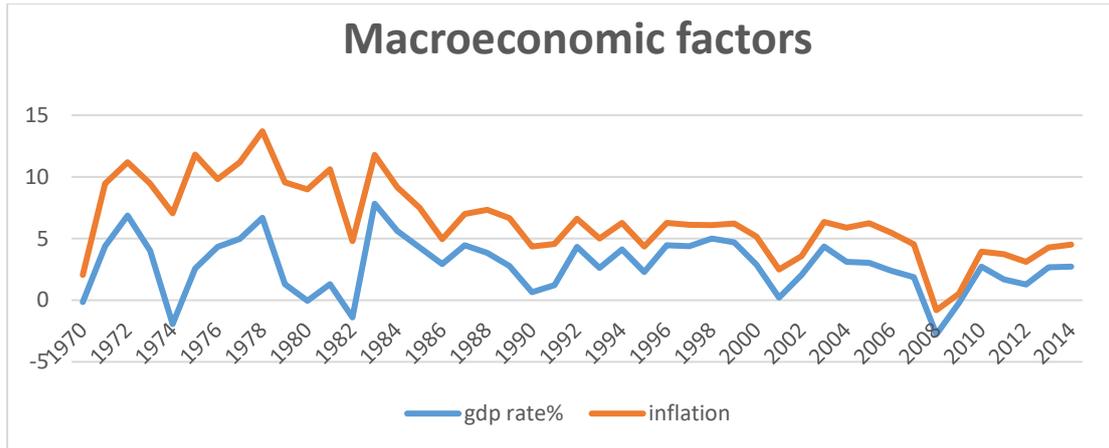

**THE ESTIMATION METHODOLOGY**

**The Quantile regression in Panel Data approach**

Our approach is based on quantile regressions, which estimate the effect of explanatory variables on the dependent variable at different points of the dependent variable's conditional distribution. Quantile regressions were originally presented as a 'robust' regression method which permits for estimation where the typical hypothesis of normality of the error term might not be strictly satisfied (Koenker and Bassett, 1978);

This method has also been used to estimate models with censoring (Powell, 1984, 1986; Buchinsky, 1994, 1995).

Recently, quantile regressions have been used simply to get evidence about points in the distribution of the dependent variable further than the conditional mean (Buchinsky, 1994, 1995; Eide and Showalter, 1997). We use quantile regressions to observe whether the effects of factors Is differentiated across the 'quantiles' in the conditional distribution of dependent variable. As described by Koenker and Bassett (1978), the estimation is done by minimizing

$$\min_{\beta \in R^K} \sum_{t \in \{t: y_t \geq x_t\beta\}} \theta |y_t - x_t\beta| + \sum_{t \in \{t: y_t < x_t\beta\}} (1-\theta)|y_t - x_t\beta|$$

where $y_t$, is the dependent variable, $x_t$ is the k by 1 vector of explanatory variables, $\beta$ is the coefficient vector and $\Theta$ is the estimated quantile. The coefficient vector b will differ depending on the particular quantile being estimated.

**RESULTS**

We choose to use the conditional quantile estimator over the OLS estimator which focuses only on the central tendency of the distribution and does not permit the possibility of differentiation of the explanatory variables impact for highly leveraged firms. An additional reason for choosing the conditional quantile estimator is that it provides us the possibility to observe the sign and probability fluctuations across quantiles, from the lower leveraged firms to the higher leveraged firms.

Table 1. Mean variables

|  | LEVB | NDTSTA | GROWTHT | INVTTA | PROFITABILITY | SIZETOTALASSETS | LIQTA | MBRATIO |
|---|---|---|---|---|---|---|---|---|
| FYEAR | Mean | | | | | | | |
| 1970 | 0.139690 | 0.042600 | 0.059778 | 0.244027 | -0.000260 | 0.132120 | 0.352779 | 1.535749 |
| 1971 | 0.141780 | 0.046729 | 0.101896 | 0.240493 | 0.004645 | 0.127837 | 0.313398 | 1.698977 |
| 1972 | 0.140992 | 0.062088 | 0.129515 | 0.242241 | 0.005716 | 0.117787 | 0.250915 | 1.678525 |
| 1973 | 0.143063 | 0.071237 | 0.076835 | 0.253919 | 0.015958 | 0.113954 | 0.344997 | 1.210098 |
| 1974 | 0.147631 | 0.024458 | -0.011965 | 0.254273 | -0.000533 | -0.003149 | 0.949054 | 0.941828 |
| 1975 | 0.150708 | 0.039875 | 0.041785 | 0.239150 | -0.025303 | -0.027838 | 0.785016 | 1.066874 |
| 1976 | 0.147880 | 0.051400 | -0.037588 | 0.236601 | -0.033367 | -0.093113 | 1.172402 | 1.148631 |
| 1977 | 0.150343 | -0.037280 | -0.240755 | 0.235098 | -0.260143 | -0.139904 | 1.182724 | 1.200959 |
| 1978 | 0.154564 | -0.105562 | 0.053530 | 0.230952 | -18.46424 | -0.368005 | 1.124757 | 1.266096 |
| 1979 | 0.153878 | 0.053043 | -0.249024 | 0.230988 | -0.308044 | -0.212858 | 1.305583 | 1.522321 |
| 1980 | 0.149778 | -0.018428 | 0.046461 | 0.218302 | -12.11516 | -0.775996 | 1.396896 | 1.978820 |
| 1981 | 0.142240 | -0.007546 | -0.048861 | 0.208034 | -16.18303 | -0.667680 | 1.365316 | 1.706371 |
| 1982 | 0.144690 | -0.070614 | -0.065758 | 0.195780 | -21.28464 | -0.192868 | 1.604093 | 1.945591 |
| 1983 | 0.133584 | -0.145602 | -0.045320 | 0.186151 | -64.82011 | -1.426612 | 1.261924 | 3.917737 |
| 1984 | 0.134885 | -0.151592 | -0.429925 | 0.190624 | -1.775894 | -0.083775 | 1.320738 | 2.127915 |
| 1985 | 0.137273 | -0.411265 | -0.093992 | 0.179714 | -220.3048 | -0.405008 | 1.790088 | 2.496286 |
| 1986 | 0.137903 | -0.245351 | -0.829047 | 0.173315 | -45.67793 | -0.780753 | 1.648475 | 4.925750 |
| 1987 | 0.140365 | -0.313044 | -0.082686 | 0.170558 | -2.311457 | -0.620398 | 1.595128 | 2.793230 |
| 1988 | 0.142941 | -0.090066 | -0.338502 | 0.171510 | -20.48788 | -1.888683 | 1.098588 | 2.021181 |
| 1989 | 0.146296 | -0.090016 | -0.607975 | 0.167783 | -13.52642 | -2.112387 | 1.392790 | 2.036429 |
| 1990 | 0.143482 | -0.456505 | -1.964750 | 0.165067 | -56.16537 | -1.679670 | 1.031316 | 1.954070 |
| 1991 | 0.137889 | 0.061399 | -0.900192 | 0.159298 | -183.4780 | -0.955596 | 1.306244 | 2.362401 |
| 1992 | 0.133088 | -0.212463 | -0.219172 | 0.156648 | -16.69850 | -1.242420 | 0.948451 | 2.367913 |
| 1993 | 0.128893 | 0.026244 | -0.036134 | 0.152011 | -85.29508 | -1.080205 | 0.874290 | 2.345391 |
| 1994 | 0.129322 | -0.183577 | -0.036700 | 0.149810 | -25.78515 | -0.111067 | 0.766826 | 2.122566 |
| 1995 | 0.131387 | -0.207771 | 0.046239 | 0.145471 | -72.40709 | -0.904528 | 0.585123 | 4.207664 |
| 1996 | 0.126173 | -0.310088 | -0.028548 | 0.139830 | -34.82157 | -0.208535 | 0.505412 | 4.230933 |
| 1997 | 0.129461 | -0.198543 | -0.328197 | 0.134392 | -12.85966 | -0.145216 | 0.506757 | 3.897723 |
| 1998 | 0.133354 | -0.393308 | -0.241248 | 0.125057 | -28.11937 | -0.300919 | 0.675374 | 3.914247 |
| 1999 | 0.133423 | -0.661480 | -0.240152 | 0.116419 | -144.4903 | -0.650102 | 1.108602 | 7.777139 |
| 2000 | 0.129199 | -0.896225 | -0.714099 | 0.113977 | -166.5361 | -2.813891 | 2.113655 | 11.63617 |
| 2001 | 0.130926 | -2.153007 | -2.235403 | 0.109943 | -533.4581 | -3.121916 | 2.117576 | 45.65303 |
| 2002 | 0.129775 | -7.194115 | -1.420282 | 0.108124 | -3945.040 | -2.407512 | 1.417264 | 18.78382 |
| 2003 | 0.127107 | -2.157808 | -0.554177 | 0.102899 | -769.3364 | -2.747282 | 1.372104 | 34.10847 |
| 2004 | 0.122018 | -2.166352 | -0.474094 | 0.100291 | -757.6105 | -2.226387 | 1.026367 | 25.38208 |
| 2005 | 0.119326 | -1.673168 | -3.844319 | 0.096758 | -498.8867 | -1.939739 | 0.941373 | 41.69120 |
| 2006 | 0.117370 | -2.294531 | -19.49168 | 0.095849 | -560.5591 | -0.620489 | 1.065100 | 12.20908 |
| 2007 | 0.117952 | -19.96244 | -0.675176 | 0.093070 | -810.5068 | -0.897278 | 1.147026 | 18.08759 |
| 2008 | 0.124018 | -5.812493 | -1.501465 | 0.096243 | -431.6909 | -1.263041 | 0.851625 | 26.51975 |
| 2009 | 0.119192 | -1.953972 | -0.306205 | 0.087959 | -537.0656 | -2.319693 | 1.425346 | 32.38875 |
| 2010 | 0.112245 | -3.175418 | -0.149307 | 0.087225 | -506.5191 | -1.897800 | 1.560945 | 33.37853 |
| 2011 | 0.114922 | -3.032119 | -0.651545 | 0.089214 | -1296.340 | -2.929817 | 1.318023 | 58.76859 |
| 2012 | 0.118043 | -2.041775 | -1.169095 | 0.082447 | -533.9204 | -4.136806 | 1.471296 | 62.41736 |
| 2013 | 0.122366 | -3.860906 | -1.184355 | 0.080221 | -980.9264 | -4.436785 | 1.711409 | 56.20476 |
| 2014 | 0.128992 | -3.245300 | -1.007030 | 0.080156 | -459.4656 | -2.251242 | 1.019701 | 44.83848 |
| All | 0.132945 | -1.582786 | | 0.149885 | -339.4475 | -1.218061 | 1.157826 | 15.33793 |

Table 1 contains the timeless process of the mean value of the variables listed in descriptive statistics table. As can be seen, the profitability shows that the businesses are not stable and do not have the capacity, even before the crisis, to produce earnings on their spending. The negative average that appears from the beginning (1970) just shows the weaknesses and not the business expenditure audited. The book value leverage fluctuated at low levels from 1970 to 1982 and 1983 to 2004 and the bottomed values noticed from 2005 to 2012. The variable market-to-book ratio seems to reflect the true value of the business in the years 2001 to 2014. After 2005 the average price of the companies rises inexplicably, warning that the market value of a company is not its relative accounting value. The size and growth of US firms remain constant during all the years except 2012, which shows a negative trend from 1974 to 2014. The liquidity variable illustrates that the asset transactions do not affect the prices, which remain unaffected. According to the table, the variable investment is only affected by the crisis from 2005 until 2014. The tax shield non-interest seems to be used by US companies to reduce their taxes owed. This happens all over the years from 2001 to 2011.

Table2.Correlation of variables

| Correlation Probability | LEVM | LEVB | NTDS_AT | GROWTHAT | INVTA | PROFTA | SIZEAT | LIQTA | MBRATIO | INFLATION | GDP_RATE |
|---|---|---|---|---|---|---|---|---|---|---|---|
| LEVM | 1.000000 ----- | | | | | | | | | | |
| LEVB | 0.774976 0.0000 | 1.000000 ----- | | | | | | | | | |
| NTDS_AT | 0.004957 0.0233 | -0.001813 0.4067 | 1.000000 ----- | | | | | | | | |
| GROWTHAT | 0.001258 0.5648 | -0.016054 0.0000 | 0.287029 0.0000 | 1.000000 ----- | | | | | | | |
| INVTA | 0.109086 0.0000 | 0.002776 0.2039 | 0.004726 0.0305 | 0.003584 0.1009 | 1.000000 ----- | | | | | | |
| PROFTA | 0.003712 0.0893 | -0.005120 0.0191 | 0.701418 0.0000 | 0.575486 0.0000 | 0.005032 0.0212 | 1.000000 ----- | | | | | |
| SIZEAT | 0.012493 0.0000 | 0.007104 0.0011 | 0.019446 0.0000 | -0.108386 0.0000 | 0.017302 0.0000 | -0.000960 0.6603 | 1.000000 ----- | | | | |
| LIQTA | -0.033672 0.0000 | -0.034693 0.0000 | -0.000571 0.7939 | -0.004025 0.0654 | -0.023213 0.0000 | -0.001450 0.5069 | -0.024153 0.0000 | 1.000000 ----- | | | |
| MBRATIO | -0.019342 0.0000 | -0.002551 0.2430 | -0.527220 0.0000 | -0.307033 0.0000 | -0.015617 0.0000 | -0.378144 0.0000 | -0.364534 0.0000 | 0.010543 0.0000 | 1.000000 ----- | | |
| INFLATION | 0.203700 0.0000 | 0.079925 0.0000 | 0.004534 0.0379 | 0.001399 0.5220 | 0.263124 0.0000 | 0.002764 0.2057 | 0.009036 0.0000 | 0.004239 0.0523 | -0.013508 0.0000 | 1.000000 ----- | |
| GDP_RATE_ | -0.022126 0.0000 | 0.013480 0.0000 | 0.006462 0.0031 | 0.001301 0.5515 | 0.059024 0.0000 | 0.001688 0.4397 | 0.005138 0.0187 | -0.000371 0.8653 | -0.008721 0.0001 | -0.145853 0.0000 | 1.000000 ----- |

Table 2 illustrates the correlation coefficients between the variables used in our model. The dependent and independent variables are provided with a Pearson correlation matrix. It is conspicuous that there is a non-statistically significant and negative correlation (r=-0.001821) between leverage in book values and non-debt tax shields. Similar findings were obtained by Frank & Goyal (2003). A statistically significant and positive correlation between leverage and investments at the 5% significance level, a statistically negative correlation (r=-0.0328) between leverage and liquidity at the 5% significance level (Pecking order theory), a statistically significant and negative correlation (r=-0.034814) between leverage and profitability at the 5% significance level. Pecking-order theory predicts a negative relationship between profitability and leverage (accounting and market). This theory argues that companies will prefer to finance their needs first by using sustainable profits, then through borrowing and then through the issuance of new shares. According to the Pecking-order theory, companies that are rapidly developing and have high funding needs will move on to short-term funding that is less subject to asymmetric information.

Furthermore, Pearson correlation provide us information for positive relation (r=0.0071) between leverage and size. This finding is in line with trade-off theory, that is, the bigger the business is, the greater the ability to borrow, and therefore it can have a higher leverage than a smaller company. According to Titman & Wessels (1988), the bigger the business-diversity, the shorter the probability of bankruptcy will be, and the less volatility is observed in its cash flows, so it can borrow to a larger extent from smaller companies. Finally, there is a statistically significant and negative correlation (r=-0.0161) between leverage and growth at the 5% significance level. Regarding trade-off theory, companies with high growth (investment) prospects must have low levels of borrowing.

**Hausman Test**

With the regression equation, we will choose the most catalyzed model between fixed effects and random effects and with the help of the Hausman test we will make the most appropriate choice for our model.

Correlated Random Effects - Hausman Test
Equation: RNDOM_LEVB
Test cross-section random effects

| Test Summary | Chi-Sq. Statistic | Chi-Sq. d.f. | Prob. |
| --- | --- | --- | --- |
| Cross-section random | 140.152192 | 7 | 0.0000 |

$H_0$ : Random effects model is appropriate

$H_1$ : Fixed effect model is appropriate.

Probability of Chi-Sq < 0.05, so we reject null hypothesis, and Fixed Effect Model is the most appropriate for our model.

**Table 3.**

| MARKET LEVERAGE | QUANTILES | | | | |
|---|---|---|---|---|---|
| | 0,15 | 0,35 | 0,5 | 0,75 | 0,95 |
| LIQUIDITY | -0,0005 | -0.0001*** | -0.0001*** | -0.0001*** | -0.0001*** |
| sterrors | 0.0003 | 4.53E-05 | 4.95E-05 | 1.84E-05 | 1.62E-05 |
| MBRATIO | -5.59E-06 | -1.38E-05*** | -1.57E-05*** | -1.76E-05*** | -3.14E-05*** |
| sterrors | 5.03E-06 | 1.33E-06 | 8.72E-07 | 1.95E-06 | 4.77E-07 |
| NDTS | -5.16E-06 | -1.16E-05*** | -1.25E-05*** | -3.00E-05 | -5.11E-05*** |
| sterrors | 5.09E-06 | 1.37E-06 | 9.00E-07 | 0.0001 | 3.00E-06 |
| PROFITABILITY | -1.21E-08 | -3.88E-08*** | -4.89E-08*** | -3.27E-08 | -8.45E-09 |
| sterrors | 8.09E-09 | 2.76E-09 | 1.84E-09 | 3.61E-07 | 5.53E-09 |
| SIZE | -1.86E-06 | -1.08E-05*** | -9.76E-06*** | -1.60E-05*** | -2.83E-05*** |
| sterrors | 8.17E-06 | 2.31E-06 | 2.05E-06 | 3.00E-06 | 2.22E-06 |
| GROWTH | -1.68E-06 | 5.97E-06*** | 9.62E-06*** | 9.38E-06 | 2.65E-06*** |
| sterrors | 1.56E-06 | 1.61E-06 | 1.12E-06 | 6.61E-05 | 5.94E-07 |
| INVESTMENTS | -0.0083*** | -0.0145*** | -0.0031*** | 0.0238*** | 0.0676*** |
| sterrors | 0.0013 | 0.0014 | 0.0010 | 0.0016 | 0.0075 |
| INFLATION | 0.0025*** | 0.0025*** | 0.0022*** | 0.0025*** | -0.0042*** |
| sterrors | 9.85E-05 | 0.0001 | 0.0001 | 0.0001 | 0.0004 |
| GDPRATE | -0.0008*** | -0.0026*** | -0.0030*** | -0.0047*** | -0.0129*** |
| sterrors | 0.0001 | 8.77E-05 | 7.42E-05 | 0.0001 | -0.012919 |
| FIXED_EFFECTS | 0.4395*** | 0.8202*** | 0.9920*** | 1.2649*** | 1.2701*** |
| R-squared | 11.1% | 34.8% | 44.6% | 50.2% | 42% |

Assets structure variable (LAS) regarding total debt, enters with a positive sign (Bradley et al. (1984); Kaur& Rao (2009)) and finally shifts to a negative sign (Nguyen & Ramachandran (2006); Al- Ajmi et al. (2009); Karadeniz et al. (2009); Matzaz &Dusan (2009); Sheikh & Wang (2011)) from 0.6 quantile remaining statistically significant.

The fact that SIZE (SIZE) enters with an insignificantly negative coefficient and from 0.35 quantile becomes significant with the same negative sign and rises since the 0.9 quantile of market leverage ratio, indicates that larger enterprises are less diversified and can be expected to bankrupt more often while smaller firms are usually opaquer well. At this point, larger firms can be expected for lower levels of leverage.

Growth (LGR), enters with a negative sign and from 0.35 quantile becomes positive until the 0.9 quantile that is statistically significant for the total of quantiles. Firms with higher growth opportunities are more likely to exhaust internal funds and seek external financing.

Profitability (Lprof), appears with a negative sign that is non-statistically significant and from 0.35 until 0.50 quantile the sign turns to negative and significant; a fact that indicates that higher total-leveraged firms in US, according to the Trade-off theory, are less profitable while less total-leveraged firms according Pecking order theory for financing decisions follow a preference for internal over external financing and for debt over equity.

Tax considerations are very noticeable for enterprises because of the reason that they can produce high profit. Non-debt-tax-shields enter with negative and non-significant sign and remain negative and statistically significant. This suggests that the relative advantage of resorting to debt as a tax shield alternative to depreciation is lower for high levels of total leverage.

Liquidity remains negative and significant. Cash-rich companies expected to have lower debt and they prefer internal financing (Pecking order theory).

The fact that inflation (infl) starts with positive sign and from 0.95 quantile becomes negative, indicates that the high-leveraged firms are influenced from inflation in contrast to leveraged and less-leveraged firms, Frank and Goyal (2009).

**Table 4**

| BOOK LEVERAGE | QUANTILES | | | | |
|---|---|---|---|---|---|
| | 0,15 | 0,35 | 0,5 | 0,75 | 0,95 |
| LIQUIDITY | -0.0008 | -0.0001** | -0.0001*** | -0.0002*** | -0.0001*** |
| sterrors | 0.0005 | 8.95E-05 | 3.73E-05 | 1.43E-05 | 8.03E-06 |
| MBRATIO | -1.92E-05*** | -6.18E-06*** | -4.17E-06*** | -3.09E-06*** | 0.0006 |
| sterrors | 5.37E-06 | 7.46E-07 | 3.90E-07 | 7.50E-07 | 0.0004 |
| NDTS | 3.94E-05*** | -2.56E-06*** | 3.62E-07 | 2.80E-06*** | -0.0013 |
| sterrors | 1.40E-05 | 7.43E-07 | 4.29E-07 | 7.85E-07 | 0.0009 |
| PROFITABILITY | -8.13E-08*** | 7.11E-09*** | 2.71E-09*** | -1.79E-09 | 2.87E-06 |
| sterrors | 2.83E-08 | 1.40E-09 | 9.67E-10 | 2.19E-09 | 1.83E-06 |
| SIZE | 1.48E-05*** | 6.72E-06*** | 1.05E-05*** | 9.87E-06*** | 8.58E-05*** |
| sterrors | 1.72E-06 | 1.48E-06 | 1.30E-06 | 2.39E-06 | 1.67E-05 |
| GROWTH | 2.61E-05 | -4.37E-05*** | -4.09E-05*** | -3.89E-05*** | -0.0023 |
| sterrors | 3.67E-05 | 8.80E-07 | 5.05E-07 | 5.37E-07 | 0.0443 |
| INVESTMENTS | -0.0089*** | -0.0322*** | -0.0303*** | -0.0166*** | -0.0582*** |
| sterrors | 0.0011 | 0.0010 | 0.0005 | 0.0010 | 0.0094 |
| INFLATION | 0.0024*** | 0.0004*** | -0.0006*** | -0.0015*** | -0.0068*** |
| sterrors | 8.60E-05 | 7.46E-05 | 4.23E-05 | 5.92E-05 | 0.0004 |
| GDPRATE | 0.0004*** | -0.0003*** | -0.0004*** | -0.0005*** | -0.0025*** |
| sterrors | 9.02E-05 | 7.26E-05 | 3.49E-05 | 6.50E-05 | 0.0003 |
| FIXED_EFFECTS | 0.5706*** | 0.9224*** | 1.0112*** | 1.2272*** | 1.3335*** |
| R-squared | 11.5% | 38.5% | 45.6% | 44.2% | 34.2% |

Growth (LGR), enters with a positive sign and from 0.35 quantile becomes negative until the 0.95 quantile and it is statistically significant for majority of quantiles. Firms with higher growth opportunities are more likely to seek external financing.

Profitability (Lprof), appears with a negative sign for 0.15 and 0.50 quantile and it is significant from 0.15 until 0.50 quantiles; this fact indicates that higher total-leveraged firms in US according to the Trade-off theory, are less profitable. Total-leveraged firms according to Pecking order theory, regarding their financing decisions follow a preference for internal over external financing and for debt over equity.

Non-debt-tax-shields enters with positive and significant sign and remains negative and for 0.35 and 0.95 is negative. This suggests that the relative advantage of resorting to debt as a tax shield alternative to depreciation is lower for median and high levels of leverage.

The inflation (infl) starts with a positive sign and from 0.50 quantile becomes negative, which indicates that the high-leveraged firms are influenced from inflation in contrast to leveraged and less-leveraged firms, Frank and Goyal (2009).

Liquidity remains with negative and significant sign. Cash-rich companies expected to have lower debt and they prefer internal financing (Pecking order theory).

**Table 5**

|  | QUANTILES | | | | |
|---|---|---|---|---|---|
|  | 0.15 | 0.35 | 0.5 | 0.75 | 0.95 |
| SPEED MARKET | 52.7% | 59.7% | 81.2% | 74.2% | 33.2% |
| R-squared | 43.6% | 49.1% | 63.6% | 55.9% | 20.8% |
| SPEED BOOK | 25.9% | 49.3% | 53.2% | 44.4% | 7.4% |
| R-squared | 13.7% | 27.1% | 29.4% | 25.2% | 4.5% |

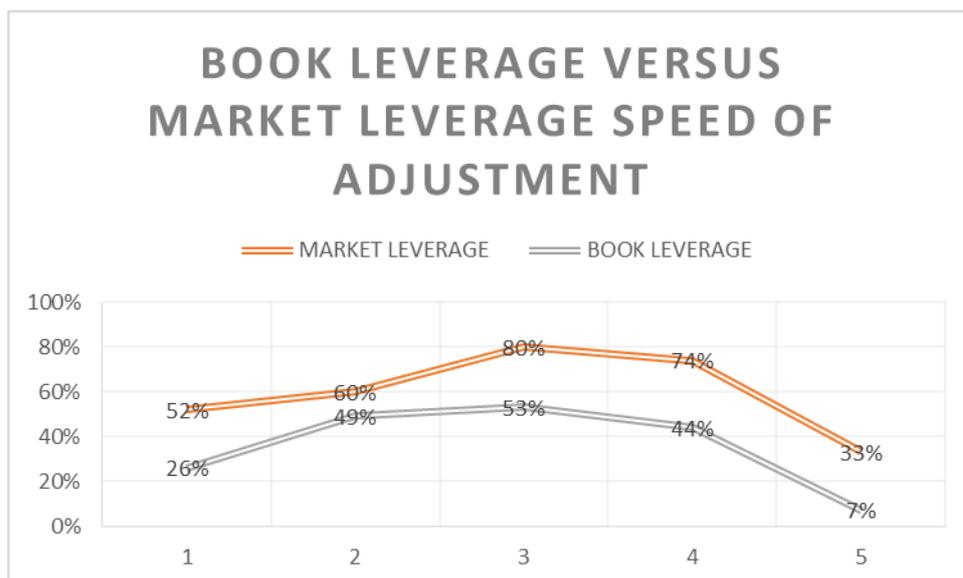

**Conclusion**

Inspired by the importance of understanding capital structure, this paper uses quantile regression approximations to add to empirical information and focuses on how capital structure relations fluctuate between firms at different stages of the total debt, short-terms debt and long-terms debt distribution. Our findings complement a new dimension to the knowledge of US firms' financing behavior reported in existing literature suggesting that research on the relation between capital structure and incentives could benefit from knowledge of heterogeneity in the capital structure and from using quantile regression techniques in the field of corporate finance. Moreover, if firms maximize subject to an upper constraint on debt, the relationship between leverage and its determinants might change sign as leverage increases, and quantile regression enables us to identify such effects within sample.

The adjustment speed slows down for long-terms debt ratio, the adjustment speed slows down only in the first quantile and from the second quantile does not change. The adjustment speed for long-terms debt ratio does not affect during crisis and the total debt ratio slows down for most of the quantiles.